\documentclass[preprint,12pt]{elsarticle}

\usepackage{graphicx}

\usepackage{amssymb}
\newcounter{bla}

\journal{Computer Physics Communications}

\begin{document}

\begin{frontmatter}

%% Title, authors and addresses

%% use the tnoteref command within \title for footnotes;
%% use the tnotetext command for the associated footnote;
%% use the fnref command within \author or \address for footnotes;
%% use the fntext command for the associated footnote;
%% use the corref command within \author for corresponding author footnotes;
%% use the cortext command for the associated footnote;
%% use the ead command for the email address,
%% and the form \ead[url] for the home page:
%%
%% \title{Title\tnoteref{label1}}
%% \tnotetext[label1]{}
%% \author{Name\corref{cor1}\fnref{label2}}
%% \ead{email address}
%% \ead[url]{home page}
%% \fntext[label2]{}
%% \cortext[cor1]{}
%% \address{Address\fnref{label3}}
%% \fntext[label3]{}

%\title{A simple variational Quantum Monte Carlo model for excitons and trions in quantum dots}
\title{A simple variational quantum Monte Carlo-effective mass approach for excitons and trions in quantum dots}

%% use optional labels to link authors explicitly to addresses:
%% \author[label1,label2]{<author name>}
%% \address[label1]{<address>}
%% \address[label2]{<address>}

\author[a]{Josep Planelles\corref{author}}
\author[a]{Juan I. Climente}

\cortext[author] {Corresponding author.\\\textit{josep.planelles@uji.es}}
\address[a]{Departament de Qu\'{\i}mica F\'{\i}sica i Anal\'{\i}tica, Universitat Jaume I, Castell\'o de la Plana, E-12080, Spain}

\begin{abstract}
A computational model is presented to calculate the ground state energy of neutral and charged excitons 
confined in semiconductor quantum dots. The model is based on the variational Quantum Monte Carlo method
and effective mass Hamiltonians. Through an iterative Newton-Rhapson process, minimizing the local energy,
and (optional) parallelization of random walkers, fast and accurate estimates of both confinement and 
Coulomb binding energies can be obtained in standard desktop computers.
To illustrate the reach of the model, we provide \emph{Fortran} programs and illustrative calculations 
for colloidal CdSe nanoplatelets with large lateral dimensions and dielectric confinement, 
where electronic correlations are strong. 
The results compare well with exact variational calculations and largely outperform configuration 
interaction calculations in computational efficiency.
\end{abstract}


\end{frontmatter}

%\linenumbers  

% Computer program descriptions should contain the following
% PROGRAM SUMMARY.

{\bf Program Summary }

\begin{small}
\noindent
{\em Program Title: vqmc-ema }                                    \\
{\em Licensing provisions: GPLv3 }                                   \\
{\em Programming language: Fortran 90}                                   \\
 
{\em Nature of problem:}\\
Calculation of either exciton or trion ground state energy and wave function in a 
cuboidal semiconductor nanoplatelet. Hard wall quantum confinement
coexists with dielectric confinement (self-energy and Coulomb polarization terms).

{\em Solution method:}\\
Variational quantum Monte Carlo with effective mass Hamiltonians, 
integrated into a Newton-Rhapson solver.
OMP parallelization library can be (optionally) linked.\\

\end{small}

%% main text
\section{Introduction}
\label{s:intro}

Coulomb interactions within excitonic species in semiconductor quantum dots (QDs)
determine optoelectronic properties of interest, such as the emission wavelength
and thermal stability (both related to the binding energy), or 
the radiative recombination rate (related to exciton Bohr radius).\cite{Pawel_book,Bimberg_book,Tapash_book}
 
Early studies in self-assembled QDs and colloidal nanocrystals showed that perturbative 
estimates of Coulomb interaction provide a good approximation of emission energies (about 95\%), 
but they miss electronic correlations, which are important in determining binding energies and 
effective Bohr radius, especially for particles beyond neutral excitons.\cite{ShumwayPRB}
Recently developed colloidal nanoplatelets (NPLs)\cite{LhuillierACR,DirollJMCc} are even more demanding.
These nanostructures are in an intermediate confinement regime between that of QDs
and that of quantum wells\cite{RichterPRM}, and present strong dielectric confinement which
enhances Coulomb interactions. In these structures, large exciton binding energies (around $200$ meV)
are observed, with non-perturbative terms representing over half that value.\cite{RajadellPRB}
An appropriate description of excitonic interactions becomes essential to study the
photo-physics.

In order to account for electron-hole and electron-electron correlation in
QDs, configuration interaction (CI) methods are arguably the most 
widely employed models to date.\cite{Pawel_book,Bimberg_book,Tapash_book,ShumwayPRB,ClimentePRB,RontaniJCP,CorniPRB,BallesterJAP}
These methods are generally built upon a basis set of independent particle or Hartree-Fock states,
which has been found to provide an excellent description of repulsions in 
few- and many-fermion systems.\cite{Pawel_book,RontaniJCP,BallesterJAP}
The description of strong attractions is however more demanding, because the afore 
mentioned basis functions are less suited to account for short-range interactions.\cite{RontaniJPB}
For the same reason, in charged excitons (trions) and multiexcitons they may not
describe attractions and repulsions with comparable accuracy. 
As a result, CI calculations aiming at precise binding energies often require
large basis sets,\cite{ShumwayPRB,ClimentePRB} which make the method computationally demanding.

Quantum Monte Carlo (QMC) methods are an alternative to CI techniques to calculate
correlations with arbitrary accuracy.\cite{CeperleyPRB,NeedsJPCM} 
Unlike CI models, QMC ones provide a description of the lowest state of each symmetry only, 
but they have the advantage of showing no pathological behavior for short-range interactions. 
Studies of excitonic complexes in QDs relying on QMC calculations have been 
reported.\cite{ShumwayPRB,HuPRL,WangJAP,WimmerPRB,LeeNL,McDonaldPRB}
Unfortunately, these methods are computationally expensive, which has prevented widespread use to date.
Continuum QMC methods, which encompass variational and diffusion QMC,\cite{NeedsJPCM} 
have been implemented in high-performance software packages such as 
CASINO\cite{casino} or QMCpack\cite{qmcpack}.
However, these programs were originally developed for molecular and
periodic crystal systems, and then take advantage of Jacobi coordinates 
in the Hamiltonian and wave function. 
This is well suited for atomistic descriptions of QDs.\cite{WilliamsonPRL}
 The description of QDs with simple and intuitive effective mass Hamiltonians, 
which have proved extremely useful to understand many electronic
properties of QDs,\cite{Pawel_book,Bimberg_book,Tapash_book}
is however handicapped by the usual presence of potentials forbidding the
use of relative coordinates. A few examples are non-parabolic confinement, 
strain and non-local interaction with self-image charges.\cite{Bimberg_book,KumagaiPRB}

%for large (quasi-periodic) QDs\cite{BauerPRB} 

In this work we present a QMC model for the study of excitons and trions
in QDs within the effective mass formalism. The goal is to provide a flexible
theoretical framework, capable of describing structures all the way from
strongly confined to weakly confined regimes, with computational requirements
affordable by ordinary desktop computers.
To this end, we use the simple yet elegant variational Quantum Monte Carlo (VQMC) method, 
which offers a ground state energy solely limited by the choice of an appropiate trial wave function.\cite{NeedsJPCM}
The latter is defined within envelope function approximation in effective mass theory.\cite{Bastard_book}
The model is integrated into a Newton-Rhapson solver to optimize the variational 
parameters in few steps.
A suit of \emph{Fortran} codes developed for the specific case of colloidal CdSe NPLs, 
which constitute a particularly demanding system owing to the strong electronic correlations, 
is provided.  Illustrative calculations show that the method largely outperforms 
CI calculations in terms of accuracy and computational efficiency, 
for both neutral excitons and trions.

\section{Theoretical framework}

\subsection{Hamiltonians and trial functions}

\subsubsection{Neutral Exciton}

Electron and hole states are described with single-band k$\cdot$p Hamiltonians.
For excitons, the Hamiltonian is:
\begin{equation}
{\hat H}_X = \sum_{i=e,h} \left( \frac{{\hat \mathbf{p}}_i\,^2}{2 m_{i}} + V_i \right) + V_c(\mathbf{r}_e,\mathbf{r}_h) + E_{gap},
\label{eqH_X}
\end{equation}
\noindent where $e$ and $h$ stand for electron and hole, $m_{i}$ is the effective mass of carrier $i$, 
${\hat \mathbf{p}}_i$ the momentum operator, $V_i$ the single-particle potential, 
$V_c(\mathbf{r}_e,\mathbf{r}_h)$ the Coulomb interaction between electron and hole, 
and $E_{gap}$ the bulk energy gap. 
The trial wave function we use for this system is:
\begin{equation}
\Psi_X(\mathbf{r}_e,\mathbf{r}_h,\sigma_e,\sigma_h) = 
	\Phi_e(\mathbf{r}_e) \,
	\Phi_h(\mathbf{r}_h) \,
	J(r_{eh})\,
	\sigma_e \, \sigma_h,
	\label{eq:PsiX}
\end{equation}
\noindent where $\Phi_{e}$ and $\Phi_{h}$ are the envelope functions of non-interacting electron and hole, 
which vary smoothly within the QD dimensions, $L$, and $\sigma_{e(h)}$ is the associated spin function.
$\Phi_e$ and $\Phi_h$ should preferrably be analytical functions, 
to allow rapid and exact evaluation of energies, gradients and Hessians, which we shall use next.
This is often possible in QDs under the envelope function formalism.\cite{Pawel_book,Tapash_book,Bastard_book}
$J(r_{eh}$) is a correlating Jastrow factor, which depends explicitly on the separation 
between electron and hole, $r_{eh}$.  We propose to use a short range Slater function:
\begin{equation}
J(r_{eh})=e^{-a r_{eh}},
\end{equation}	
\noindent where $a=\alpha/r_B^X$ is a variational coefficient, $r_B^X$ the exciton Bohr radius and $\alpha$ the 
actual variational parameter.  The choice of this factor is driven by the fact that it captures the correct limits 
of weak and strong Coulomb interaction with a single variational parameter. 
When interactions are secondary to confinement ($L \ll r_B^X$), 
$J(r_{eh}) \rightarrow 1$ and the independent particle scheme is retrieved. 
When interactions are dominant ($L \gg r_B^X$), $\Psi_X \rightarrow J(r_{eh})$,
which ensures a hydrogen atom-like function is retrieved, where carriers are bound to each other through Coulomb attraction.
In the latter case, $J(r_{eh})$ permits fulfilling the Kato cusp condition,\cite{KatoCPAM} 
whereby kinetic energy %, $-1/2 \nabla \frac{1}{\mu} \nabla \Psi$ (with $\mu$ the exciton reduced mass) 
compensates Coulomb potential at $r_{eh}\rightarrow 0$, preventing energy divergence.
It has been shown in the 2D harmonium problem that a wave function like that in Eq.~(\ref{eq:PsiX})
gives energies close to the exact solution for a broad range of confinement strengths.\cite{PlanellesTCA} \\

\subsubsection{Trion}

We consider a positively charged trion (one electron, two holes), albeit the procedure is analogous for the negatively charged one.
The Hamiltonian is:
\begin{equation}
	{\hat H}_{X^+} = \sum_{i=e,h1,h2} \left( \frac{{\hat \mathbf{p}}_i\,^2}{2 m_i} + V_i \right) 
	+ V_c(\mathbf{r}_e,\mathbf{r}_{h1}) 
	+ V_c(\mathbf{r}_e,\mathbf{r}_{h2})   
	+ V_c(\mathbf{r}_{h1},\mathbf{r}_{h2}) 
	+ E_{gap}.
\label{eqH_X+}
\end{equation}
We use a Slater-Jastrow trial wave function for the singlet ground state:
\begin{equation}
	\Psi_{X^+}(\mathbf{r}_e,\mathbf{r}_{h1}, \mathbf{r}_{h2},\sigma_e,\sigma_{h1}, \sigma_{h2}) =   
    	\Phi_e(\mathbf{r}_e) \,
	\Phi_h(\mathbf{r}_{h1},\mathbf{r}_{h2}) \,
	J(r_1,r_2,r_{12})\,
%	\sigma_e \, \left[ \alpha(\sigma_{h1}) \beta(\sigma_{h2}) - \beta(\sigma_{h1}) \alpha(\sigma_{h2} \right].
	\sigma_e \, \left[ \alpha_{h1} \beta_{h2} - \beta_{h1} \alpha_{h2} \right],
	\label{eq:PsiX+}
\end{equation}
\noindent and choose the following Jastrow factor:
\begin{equation}
J(r_1,r_2,r_{12}) = e^{-Z r_1} \, e^{-Z r_2} \, e^{b \, r_{12}/(1+a \, r_{12})}.
\end{equation}
\noindent where $r_{1(2)}=|\mathbf{r}_e - \mathbf{r}_{h1(2)}|$ 
and $r_{12}=|\mathbf{r}_{h1}-\mathbf{r}_{h2}|$.
The first two terms are short range cusp forms describing the correlation of each hole 
with the electron, as in the exciton case. $Z=\zeta/r_B^X$ is a variational coefficient,
with $\zeta$ the parameter to be varied. 
The last term is a Pad\'e Jastrow factor, 
which has the property of giving the desired limits with $r_{12}$.
At short ranges of interaction, $r_{12}\rightarrow 0$, the term becomes $e^{b\,r_{12}}$, 
which provides a cusp to compensate for the divergence in hole-hole Coulomb repulsion ($b>0$).
At the same time, the probability to find distant holes ($r_{12} \rightarrow \infty$)
is more likely than that of proximal holes ($r_{12} \rightarrow 0$) by a factor $(e^{b/a})^2$. 
 We define $b=\beta/r_B^X$ and $a=\alpha/r_B^X$,
and let $\beta$ and $\alpha$ be the variational parameters. % tmp *** update codes with r_B^X instead of r_B^h
More sophisticated trial functions have been suggested for trions in QDs
or wells (see e.g. Ref.\cite{SergeevPSS}), but the present proposal has the advantage 
of keeping the smallest number of parameters that captures the correct limit behavior,
while being physically consistent with that of neutral excitons, Eq.~(\ref{eq:PsiX}).
%Notice that if hole-hole repulsion is weak ($b \rightarrow 0$), Eq.~(\ref{eq:PsiX+}) 
%gives the wave function of two excitons as described by Eq.~(\ref{eq:PsiX}).

\subsection{Variational Monte Carlo}

Within the VQMC framework,\cite{CeperleyPRB,NeedsJPCM} the variational energy 
corresponding to a Hamiltonian ${\hat H}$ (in our case ${\hat H}_X$ or ${\hat H}_{X+}$) is:
\begin{equation}
	\langle E \rangle = 
%	\frac{\int d\mathbf{R}\,\Psi_X^*(\mathbf{R}) {\hat H}_X \Psi_X(\mathbf{R}) }
%	{\int d\mathbf{R'}\,\Psi_X^*(\mathbf{R'}) \Psi_X(\mathbf{R'})}
	\frac{\int d\mathbf{R}\,\Psi^*(\mathbf{R}) {\hat H} \Psi(\mathbf{R}) }
	{\int d\mathbf{R'}\,\Psi^*(\mathbf{R'}) \Psi(\mathbf{R'})}
	=	
	\int d\mathbf{R}\, p(\mathbf{R}) E_L(\mathbf{R}), 
	\label{eq:Ev}
\end{equation}
\noindent where $\Psi(\mathbf{R})$ is the trial function, $E_L$ is the local energy:
\begin{equation}
%	E_L(\mathbf{R}) = \frac{{\hat H}_X \Psi_X (\mathbf{R}) }{\Psi_X(\mathbf{R})}.
	E_L(\mathbf{R}) = \frac{{\hat H} \Psi (\mathbf{R}) }{\Psi(\mathbf{R})}.
	\label{eq:El}
\end{equation}
\noindent and $p(\mathbf{R})$ the probability distribution:
\begin{equation}
p(\mathbf{R})=|\Psi(\mathbf{R})|^2/\int d\mathbf{R'} |\Psi(\mathbf{R'})|^2.
\label{eq:p}
\end{equation}
The Metropolis algorithm is used for importance sampling.
Thus, particles are moved to different trial positions in a random walk. 
The new positions are accepted if the following condition is satisfied:
\begin{equation}
%P(\mathbf{R}_{new} \leftarrow \mathbf{R}_{old}) = min[1,|\Psi(\mathbf{R}_{new}) / \Psi(\mathbf{R_{old})}|^2].
\frac{p(\mathbf{R}_{new})}{p(\mathbf{R}_{old})} > w, 
%P(\mathbf{R}_{new} \leftarrow \mathbf{R}_{old}) = min[1,|\Psi(\mathbf{R}_{new}) / \Psi(\mathbf{R_{old})}|^2].
\label{eq:accept}
\end{equation}
\noindent where $w$ is a random number homogeneously distributed between 0 and 1.
For a sufficiently long calculation, this random walk allows
to simplify Eq.~(\ref{eq:Ev}) as:
\begin{equation}
	\langle E \rangle \approx 
	\frac{1}{N}\,\sum_{i=1}^{N} E_L (\mathbf{R}_i),
	\label{eq:std}
\end{equation}
\noindent where $N$ is the number of (accepted) points $R_i$ taken in the walk.
Alternatively, an equivalent form can be used in the estimator, 
which considers not only accepted steps, but also rejected ones:
\begin{equation}
	\langle E \rangle \approx 
	\frac{1}{N}\,\sum_{i=1}^{N} 
%	\left[ P(\mathbf{R}_{new} \leftarrow \mathbf{R}_{old})\, E_L (\mathbf{R}_{new}) 
%	+ (1 - P(\mathbf{R}_{new} \leftarrow \mathbf{R}_{old}))\, E_L (\mathbf{R}_{old}) \right]
	\left[ p(\mathbf{R}_{new})\, E_L (\mathbf{R}_{new}) + (1 - p(\mathbf{R}_{new}))\, E_L (\mathbf{R}_{old}) \right]
	      \label{eq:avg}
\end{equation}
\noindent with $N$ here being the total number of attempted moves (whether accepted or not).
Eqs.~(\ref{eq:std}) and (\ref{eq:avg}) have the same average, but the latter reduces the 
fluctuations caused by the acceptance of unlikely configurations and tends to have smaller variance.\cite{CeperleyPRB}
As a result, for all the cases under study in this work, Eq.~(\ref{eq:avg}) is found to 
be at least as accurate as Eq.~(\ref{eq:std}) in obtaining the correct energy 
$\langle E \rangle$ with the same number of points $N$, and often more so. 
We shall then opt for it.

When evaluating the local energy, $E_L$, it is convenient to rewrite the 
kinetic energy term of a particle $i$, 
\begin{equation}
	K_i = -\hbar^2/(2 m_i)\, (\nabla^2 \Psi)/\Psi, 
\end{equation}
\noindent as 
\begin{equation}
K_i = 	2\,T_i - F_i^2, 
\label{eq:TF}
\end{equation}
\noindent where:
\begin{equation}
T_i = -\frac{1}{2}\,\frac{\hbar^2}{2 m_i}\, \nabla_i^2\ \ln{\Psi},
\end{equation}
\noindent and:
\begin{equation}
F_i = \left(\frac{\hbar^2}{2 m_i}\right)^{1/2}\, \nabla_i\ \ln{\Psi}.
\end{equation}
Both forms of the kinetic energy term give the same average values, 
but latter has the advantage that the derivation of logarithms provides simpler mathematical 
expressions when the wave function is composed by products of functions, 
as in $\Psi_X$ and $\Psi_{X^+}$.
It can be further shown that $T_i = F_i^2$, which is sometimes used to simplify
$2\, T_i - F_i^2 = T_i$. However, Eq.~(\ref{eq:TF}) gives significantly 
smaller variance.\cite{CeperleyPRB} 
%We then calculate kinetic energies with 
%$2\,{\hat T}_i - {\hat F}_i^2$.

\subsubsection{Random walk in confined structures}

A brief discussion on the random walk nature is in order.
In the Metropolis algorithm, trial steps in the walk are taken as a change in position
$\mathbf{R}_{new}=\mathbf{R}_{old} + \mathbf{r}$, where $\mathbf{r}$ is a random vector uniformly 
distributed in a cube of predefined side $r_0$ centered at $\mathbf{R}_{old}$. 
Random walks are usually applied to particles one by one.\cite{CeperleyPRB}

Hamiltonians (\ref{eqH_X}) and (\ref{eqH_X+}) are written in absolute coordinates,
because the single-particle potentials $V_i$ generally prevent the use of Jacobi coordinates.
On the other hand, excitonic complexes in QDs feel an attractive Coulomb interaction 
$V_c(\mathbf{r}_e,\mathbf{r}_h)$, which shows a sharp peak as electron and hole approach coalescence
($r_{eh} \rightarrow 0$). This short-range potential can have a profound impact on the wave function, 
especially if confinement is weak.
In order to obtain reliable importance sampling,
we observe that local energy and other magnitudes of interest can be evaluated with absolute, 
Cartesian coordinates, but the random walk must be applied on the corresponding relative coordinates. 
Once the new position is set, we revert to absolute coordinates again.
For the neutral exciton, relative coordinates can be center-of-mass and 
relative electron-hole motion.
For the positive trion, optimal sampling is obtained using absolute coordinates
for the electron, and relative coordinates of the two holes with respect to the electron.
By doing so, the relative motion provides the Metropolis algorithm sensibility to 
the Coulomb singularity near $r_{eh}=0$, while center-of-mass 
(or absolute) motion provides sensibility to the confinement of the QD walls.
Because of the different lengthscale of the two potential terms, 
it is also recommended that $r_0^{rel} < r_0^{CM}$, i.e. 
the cube where $\mathbf{r}^{rel}$ is inscribed has smaller size than that of $\mathbf{r}^{CM}$, 

%In general, we have a limited knowledge on the suitable initial geometry so that
%we may guess it wrongly. We may need a certain number of Monte Carlo cycles
%before we reach the region of equilibrium configuration. When this equilibrium is
%established, the mean energy oscillates around its mean value. We should then
%discard the contributions to expectation energy value till we have reached the equilibrium.

Optimizing the maximum step size ($r_0$) in the random walk is also convenient. 
 %we include a thermalization process prior to the actual evaluation of local magnitudes.
If it is too large, only a small fraction of trial points is accepted,
and the sampling is inefficient. If it is too small, a large number of trial points
is accepted, but it takes too long to sample over the whole interval of interest.
Ideally, we want about half of the steps to be accepted.\cite{CeperleyPRB}
Then, at the beginning of the random walk, a thermalization process should be included.
An arbitrary $r_0$ is chosen, and a random walk is taken to estimate
the ratio of accepted points (with no need to calculate local energies).
Once the walk is over, we count assets and failures and redefine the step 
size to fit the desired factor 1/2 of successful trial steps.

\subsection{Newton-Rhapson solver}

Having defined the Hamiltonians, trial wave functions and local energy estimator,
we can evaluate the average energy for a given set of variational parameters, 
$\mathbf{M} = (\alpha)$ in the case of the exciton, 
or $\mathbf{M} = (\zeta,\,\beta,\,\alpha )$ in the case of the trion.
In order to find the parameters that minimize $\langle E (\mathbf{M}) \rangle$, 
we resort to the iterative Newton-Rhapson method. 
We look for the zero of the energy gradient, $|g (\mathbf{M}) \rangle$. 
Given a set of parameters $\mathbf{M}_i$, 
the next set of parameters is given by:
\begin{equation}
	\mathbf{M}_{i+1} \approx \mathbf{M}_i - (\mathbb{H}(\mathbf{M}_i))^{-1}\,|g (\mathbf{M}_i) \rangle.
	\label{eq:NR}
\end{equation}
\noindent where $\mathbb{H}$ is the energy Hessian matrix. 
%$\mathbb{H}_{i,j}(\mathbf{m})=\frac{\partial^2 E(\mathbf{m})}{\partial m_i \partial m_j}$. %, with $i$ and $j$ being two variational parameters. 
%
\noindent After labelling $\Psi'_{M_i}$ the logarithmic derivative of $\Psi$ with respect to the variational parameter $M_i$, 
the components of $|g\rangle$ and $\mathbb{H}$ can be written as:\cite{LinJCP}
\begin{equation}
\label{eq2}                                    
	|g\rangle_{M_i} =  \frac{\partial E(\mathbf{M})}{\partial M_i} =  
	2\left[\langle E_L \Psi'_{M_i}\rangle -\langle E_L \rangle \langle \Psi'_{M_i}\rangle \right]
\end{equation}
\noindent and 
\begin{eqnarray}
 \mathbb{H}_{M_i,M_j}&=& \frac{\partial^2 E(\mathbf{M})}{\partial M_i \partial M_j} = 
	2\left[ \langle E_L \Psi''_{M_i,M_j}\rangle-\langle E_L \rangle \langle \Psi''_{M_i,M_j}\rangle \right. + \nonumber \\
	& &     2\left[\langle E_L \Psi'_{M_i} \Psi'_{M_j} \rangle- \langle E_L \rangle\langle \Psi'_{M_i} \Psi'_{M_j} \rangle \right] \nonumber \\
	& & \left. -\langle \Psi'_{M_i} \rangle \, |g\rangle_{M_j}
		   -\langle \Psi'_{M_j} \rangle \, |g\rangle_{M_i} + \langle \Psi'_{M_j} \frac{\partial E_{L}}{\partial M_i} \rangle \right].
\label{eq:2nd}
\end{eqnarray}
\noindent It is worth noting that for $i \neq j$ the last term,  $\langle  \Psi'_{M_j}\,\frac{\partial E_L}{\partial M_i} \rangle$, is not symmetric when approximated by a finite sample, whereas the true Hessian should be. 
We make it symmetric by replacing the term as %:\cite{UmrigarPRL} 
%
%\begin{equation}
$2\langle  \Psi'_{M_j}\,\frac{\partial E_L}{\partial M_i} \rangle = \langle \Psi'_{M_j} \frac{\partial E_{L}}{\partial M_i} \rangle+ \langle \Psi'_{M_i} \frac{\partial E_{L}}{\partial M_j} \rangle.$\cite{UmrigarPRL}
%\label{eq:symm}
%\end{equation}
%
%\noindent 
 Additional considerations are in order to reduce the variance in the stochastic calcualtion.
For real wave functions, the expectation value of the first derivative of the local value of any Hermitian 
operator with respect to a real parameter is zero.\cite{LinJCP} 
Thus,  $\langle \frac{\partial E_{L}}{\partial M_i} \rangle=0$. 
It is then convenient to rewrite the non-symmetric terms in covariance form,
i.e., $\langle a \, b \rangle -\langle a  \rangle  \langle b \rangle$ instead of  
$\langle a \, b \rangle$, for the fluctuations of $\langle a \, b \rangle -\langle a  \rangle  \langle b \rangle$ 
are smaller than those of  $\langle a \, b \rangle$.\cite{UmrigarPRL}
Therefore, we use:
\begin{equation}
2\langle  \Psi'_{M_j}\,\frac{\partial E_L}{\partial M_i} \rangle = 
\langle \Psi'_{M_j} \frac{\partial E_{L}}{\partial M_i} \rangle
 - \langle \Psi'_{M_j} \rangle \, \langle \frac{\partial E_{L}}{\partial M_i} \rangle
	+ \langle \Psi'_{M_i} \frac{\partial E_{L}}{\partial M_j} \rangle. 
 - \langle \Psi'_{M_i} \rangle \, \langle \frac{\partial E_{L}}{\partial M_j} \rangle. 
\label{eq:symm2}
\end{equation}
One can also note that the second derivatives of logarithmic functions vanish in some cases, 
which simplifies the results of Eq.~(\ref{eq:2nd}).
From Eq.~(\ref{eq:PsiX}), $(\Psi_X'')_{a,a}=0$. From Eq.~(\ref{eq:PsiX+}), 
$(\Psi_{X^+}'')_{Z,Z}=(\Psi_{X^+}'')_{b,b}=(\Psi_{X^+}'')_{Z,a}=(\Psi_{X^+}'')_{Z,b}=0$.

Eq.~(\ref{eq:NR}) is recursively applied until convergence.
Convergence can be set by a threshold either in the energy ($\langle E (\mathbf{M}_{i+1}) \rangle - \langle E (\mathbf{M}_i) \rangle $) 
or in the distance between variational parameters ($|\mathbf{M}_{i+1} - \mathbf{M}_i|$).\\

\subsection{Computational considerations}

\begin{figure}[h]
\begin{center}
\includegraphics[width=7cm]{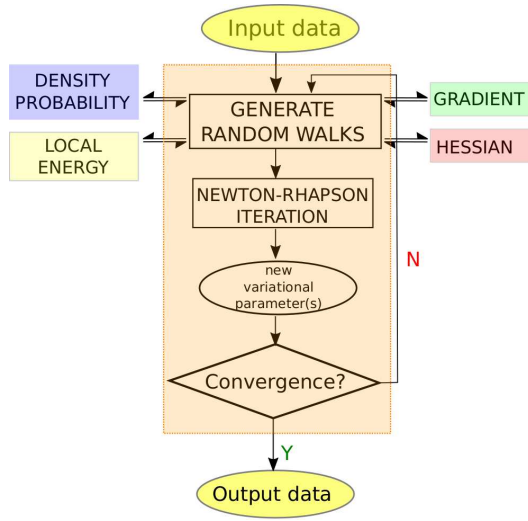}
\caption{Flowchart of the VQMC-Newton Rhapson model.}\label{fig1}
\end{center}
\end{figure}

The whole calculation method we have described can be implemented 
computationally following the flowchart of Figure \ref{fig1}.
Input data is first provided with all relevant Hamiltonian parameters
(effective masses, QD dimensions, band offsets and dielectric constants
definint potential terms) along with an initial guess of variational 
parameters ($\mathbf{M}_0$).
Random walkers are then generated to evaluate the magnitudes of interest.
In order to avoid a possible bias induced by the random initial position of a walk, 
it is customary to evaluate not one but several walkers. 
The mean value of the computed magnitude (e.g. $\langle E \rangle$) is then taken. 
From a computational point of view, it is both simple and efficient to distribute 
different walkers among different CPU processes. Rather than resorting to 
MPI parallelization, which is best suited for high-performance computing centers 
with multiple nodes/servers/computers instances, one can use OpenMP parallelization,
which is easily applied to current standard computers.

For every trial point $\mathbf{R}$, the walkers compute local magnitudes: 
density probability, energy, gradient vector and Hessian matrix.  
If the point fulfills Eq.~(\ref{eq:accept}), the step is accepted as a starting 
point for future moves.  With a sufficiently high number of points and walkers, 
average estimates of the variational values of $E(\mathbf{M}_0)$, 
$|g(\mathbf{M_0})\rangle$ and $\mathbb{H}(\mathbf{M_0})$ are obtained using 
Eq.~(\ref{eq:avg}).\cite{general,ders}
These values are used to propose the next set of variational parameters  $\mathbf{M}_1$
in a Newton-Rhapson iteration, Eq.~(\ref{eq:NR}). The process is repeated 
until convergence is achieved. The resulting energy and variational 
parameters are then written in an output file.

Because local magnitudes are calculated on-the-fly and only accumulated values are stored, 
memory requirements in the VQMC calculations we propose are minimal. 
Together with the conceptual simplicity and the ease of parallelization, 
this is an additional asset for the model to be efficient in ordinary computers.

\section{Case study: semiconductor nanoplatelets}

Semiconductor NPLs are a particularly challenging system to calculate excitonic interactions. 
These nanostructures typically show rectangular cuboid shape, with only a few atomic monolayers 
thickness ($1-2$ nm) in one direction, and tens of nm in the other two.\cite{LhuillierACR,DirollJMCc} 
They constitute intermediate structures betweeen zero-dimensional QDs and two-dimensional quantum wells.\cite{RichterPRM}
Besides, Coulomb interactions are strongly enhanced by dielectric mismatch with the surrounding organic ligands.\cite{AchtsteinNL}
As a result, correlations play a significant role and quantitative estimates of exciton properties are 
beyond perturbational and standard CI calculations.\cite{PlanellesTCA}

For neutral excitons in CdSe NPLs, an accurate description of the ground state has been obtained through a full
variational calculation of $\Psi_X$, with a number of approximations to reduce the six-dimensional Coulomb
integrals to two-dimensional ones.\cite{RajadellPRB} 
For the trion, however, the larger number of variational parameters and particle coordinates renders
this approach computationally impractical. 
In this section, we show how the VQMC model presented above enables such a study
with modest computational requirements. %available to most researchers. 

\subsection{Neutral exciton}

Following Ref.~\cite{RajadellPRB}, excitons in CdSe NPLs can be described by giving $H_X$ the specific form:
\begin{equation}
H_X = \sum_{i=e,h} \left( \frac{\mathbf{p}_\parallel\,^2}{2 m_{\parallel,i}} + \frac{p_z^2}{2 m_{z,i}} + V_i \right) + V_c(\mathbf{r}_e,\mathbf{r}_h) + E_{gap},
\label{eqH_NPL}
\end{equation}
\noindent where $m_{\parallel,i}$ is the in-plane mass of carrier $i$ and $m_{z,i}$ that along the strongly confined direction ([001]).
The single particle potential is written as $V_i = V_i^{pot} + V_i^{self}$,  where $V_i^{pot}$ is the confining potential set by the 
band offset between the NPL core and its surrounding material. 
We take $V_i^{pot}=0$ inside the NPL and $V_i^{pot}=\infty$ outside.
$V_i^{self}$ is the self-energy potential resulting from the interaction of carriers with their image charges 
in the dielectric medium, which we model as in quantum wells.:\cite{KumagaiPRB}
\begin{equation}
V_i^{self} = \sum_{n=\pm 1,\pm 2,\ldots} \frac{ q_n \, q_e^2}{2 \epsilon_{in} \left[ z_i - (-1)^n z_i + n L_z \right]}.
\end{equation}
\noindent with $q_e$ being the electron charge, $q_n=((\epsilon_{in}-\epsilon_{out})/(\epsilon_{in} + \epsilon_{out}))^{|n|}$, 
$\epsilon_{in}$ ($\epsilon_{out}$) the dielectric constants inside (outside) the NPL and $L_z$ the thickness of the NPL.

For the electron-hole Coulomb interaction, we also take into account the influence of polarization due to the dielectric mismatch:\cite{KumagaiPRB}
\begin{equation}
V_c = \sum_{n=-\infty}^{\infty} \frac{ q_n \, q_e^2 } { \epsilon_{in} \left[ (\mathbf{r}_{\parallel,e}-\mathbf{r}_{\parallel,h})^2  + \left[z_e - (-1)^n z_h + n\,L_z \right]^2 \right]^{1/2}}.
\label{eqVc}
\end{equation}

The trial wave function is of the form: 
\begin{equation}
\label{eq:PsiX_NPL}
\Psi_X(\mathbf{r}_e,\mathbf{r}_h,\sigma_e,\sigma_h) = 
	\Phi_e(\mathbf{r}_e) \,
	\Phi_h(\mathbf{r}_h) \,
%	e^{-a \sqrt{(\mathbf{r}_{\parallel,e}-\mathbf{r}_{\parallel,h})^2}} \,
	e^{-a r_{\parallel,eh}}  \,
	\sigma_e \, \sigma_h,
\end{equation}
\noindent where $r_{\parallel,eh}=\sqrt{(\mathbf{r}_{\parallel,e}-\mathbf{r}_{\parallel,h})^2}$ is
the in-plane distance between electron and hole. $\Phi_e$ and $\Phi_h$ are particle-in-box electron and hole single-particle states:
\begin{equation}
\Phi_i = \cos{k_x x_i}\, \cos{k_y y_i}\, \cos{k_z z_i}.
\label{eqPhi}
\end{equation}
Here $k_j=\pi/L_j$,  with $L_j$ being the dimensions of the NPL along the direction $j=x,y,z$. 
Notice that, because $L_z \ll L_x \sim L_y$, the Jastrow factor in Eq.~(\ref{eq:PsiX_NPL})
contains in-plane Coulomb terms only. Also, normalization constants are omitted in $\Psi_X$ because they
cancel out in the evaluation of local magnitudes, see Eq.~(\ref{eq:El}).

The program we provide in this work, \emph{vqmc-ema}, uses 
the VQMC-Newton Rhapson model to obtain the exciton ground state of Eq.~(\ref{eqH_NPL}),
with the trial function in Eq.~(\ref{eq:PsiX_NPL}).
The code needs no external libraries other than OpenMP, and only in case we want the calculation of random walkers to be paralellized.
The program determines the optimal variational parameter $\alpha$ and its associated exciton energy, 
for any input NPL dimensions and material parameters (effective masses, dielectric constants).\cite{onself}

\begin{figure}[h]
\begin{center}
\includegraphics[width=7cm]{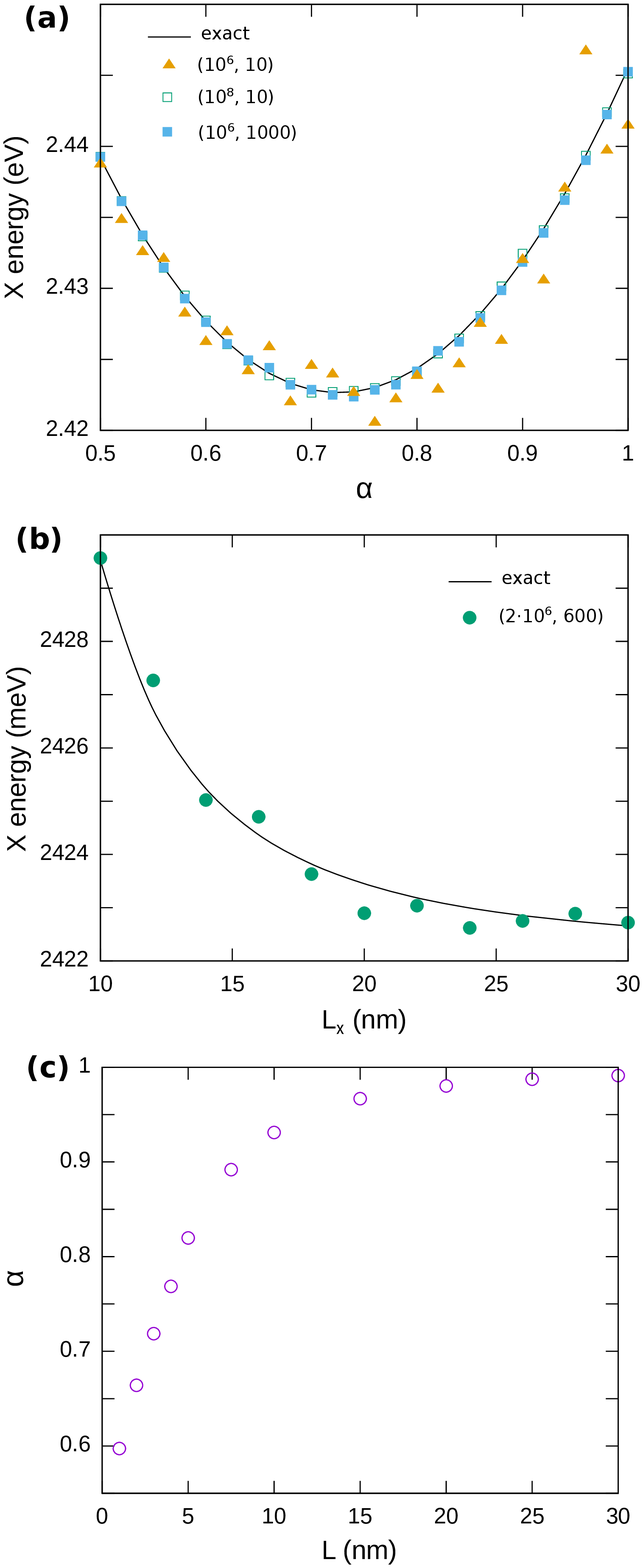}
\caption{Exciton energy in CdSe NPLs. 
(a) Energy as a function of the variational parameter $\alpha$, for a NPL
with ($L_x,L_y,L_z$)=($30,10,1.4$) nm. 
(b) Variational energy as a function of the NPL side $L_x$.
(c) Optimal variational parameter for different lateral dimensions
in a square NPL, see text for details of this calculation.
The legend is ($m,nw$), where $m$ is the number of points per walker
and $nw$ the number of walkers. 
Solid lines show the result of Ref.~\cite{RajadellPRB} model for comparison.}\label{fig2}
\end{center}
\end{figure}

A number of calculations in CdSe NPLs are run to validate the model. 
The same effective masses as in Ref.~\cite{RajadellPRB} are taken, relative dielectric constants 
are set to $\epsilon_{in}=6$ and $\epsilon_{out}=2$, and $E_{gap}=1.76$ eV.\cite{Sadao_book}
We first calculate the energy as a function of the variational parameter $\alpha$ in a NPL with 
dimensions ($L_x,L_y,L_z$)=($30,10,1.4$) nm. The result is shown in Figure \ref{fig2}(a), for different 
values of ($m, nw$), where $m$ is the number of (accepted) points per walker and $nw$ the number of walkers
used to evaluate Eq.~(\ref{eq:avg}), with $m \cdot nw = N$. The black solid line shows the result obtained
with the full variational integration of Ref.~\cite{RajadellPRB}, for comparison.
A clear minimum of $\langle E \rangle$ develops around $\alpha = 0.72$ in all cases.
Since we define the exciton Bohr radius in the 2D limit, $r_B^X=\epsilon_{in}/2\mu$ (atomic units,
 $\mu$ is the exciton in-plane reduced mass), this means the effective Bohr radius ($1/a$) in a 
NPL is intermediate between that of a 2D exciton ($\alpha=1$) and a that of 3D one ($\alpha=0.5$).
The agreement with the exact variational calculation is rough for $(m,nw)=(10^6,10)$.
However, it becomes excellent when the total number of points is increased from 
$N=10^7$ to $N=10^9$, through either larger $m$ or $nw$.

In Fig.~\ref{fig2}(b) we compare the VQMC result with the exact variational one
for different NPL sizes. Since $L_x=30$ nm, the platelet changes from rectangular 
to square with increasign $L_y$. Also, by weakening confinement, the electronic
correlations become more important. One can see in the figure that the stochastic
calculation matches the exact result in all cases with sub-meV error.
This is under $0.15\%$ relative error with respect to the total exciton 
(confinement plus Coulomb) energy.

Because our model is intended to describe excitonic interactions even
in the limit of weak confinement, we test its accuracy in fulfilling Kato cusp conditions.\cite{KatoCPAM}
To this end, because the Jastrow factor considers in plane correlations only, 
we rewrite the Coulomb terms in ($\ref{eqVc}$) removing $z$ coordinates. 
From Hamiltonian (\ref{eqH_X}) and the trial function (\ref{eq:PsiX}),
assuming homogeneous relative dielectric constant $\epsilon_{in} = \epsilon_{out}$
and negligible confinement, it is easy to check that the Coulomb divergence 
at $r_{\parallel,eh} \rightarrow 0$ is compensated by kinetic energy if $a=2\mu/\epsilon_{in}=1/r_B^X$. 
That is, a 2D hydrogen atom-like limit should be retrieved.
Fig.~\ref{fig2}(c) shows the VQMC model successfully converges towards this value
for large NPLs, since the value of $\alpha$ minimizing the exciton energy tends to 1,
and then $a \rightarrow 1/r_B^X$.  For smaller NPLs, however, kinetic energy terms coming 
from $\Phi_e$ and $\Phi_h$ make $\alpha$ deviate from the 2D limit.
The latter result illustrates that quantum confinement prevents inferring the 
value of variational parameters from simple bulk cusp conditions, and in general 
we need to optimize all the parameters in the trial wave function numerically.

\subsection{Trion}

For a positive trion in a NPL, Hamiltonian (\ref{eqH_X+}) takes the form: 
\begin{equation}
H_{X^+} = \sum_{i=e,h1,h2} \left( \frac{\mathbf{p}_\parallel\,^2}{2 m_{\parallel,i}} + \frac{p_z^2}{2 m_{z,i}} + V_i \right) 
	+ V_c(\mathbf{r}_e,\mathbf{r}_{h1}) + V_c(\mathbf{r}_e,\mathbf{r}_{h2}) + V_c(\mathbf{r}_{h1},\mathbf{r}_{h2}) 
	+ E_{gap}.
\label{eqH+_NPL}
\end{equation}
\noindent By analogy with the exciton case, for the trial wave function we use $\Psi_{X^+}$ as in
Eq.~(\ref{eq:PsiX+}) but with Jastrow factors accounting for in-plane coordinates only.
$\Phi_e$, $\Phi_{h1}$ and $\Phi_{h2}$ are particle-in-box functions, as in Eq.~(\ref{eqPhi}).
\emph{vqmc-ema} addressed the trion case as well. It optimizes the variational parameters 
$(\zeta,\beta,\alpha)$ of the ground state and provides the associated energy, 
for a given input of NPL dimensions, effective masses and dielectric constants.
\begin{figure}[t]
\begin{center}
\includegraphics[width=14cm]{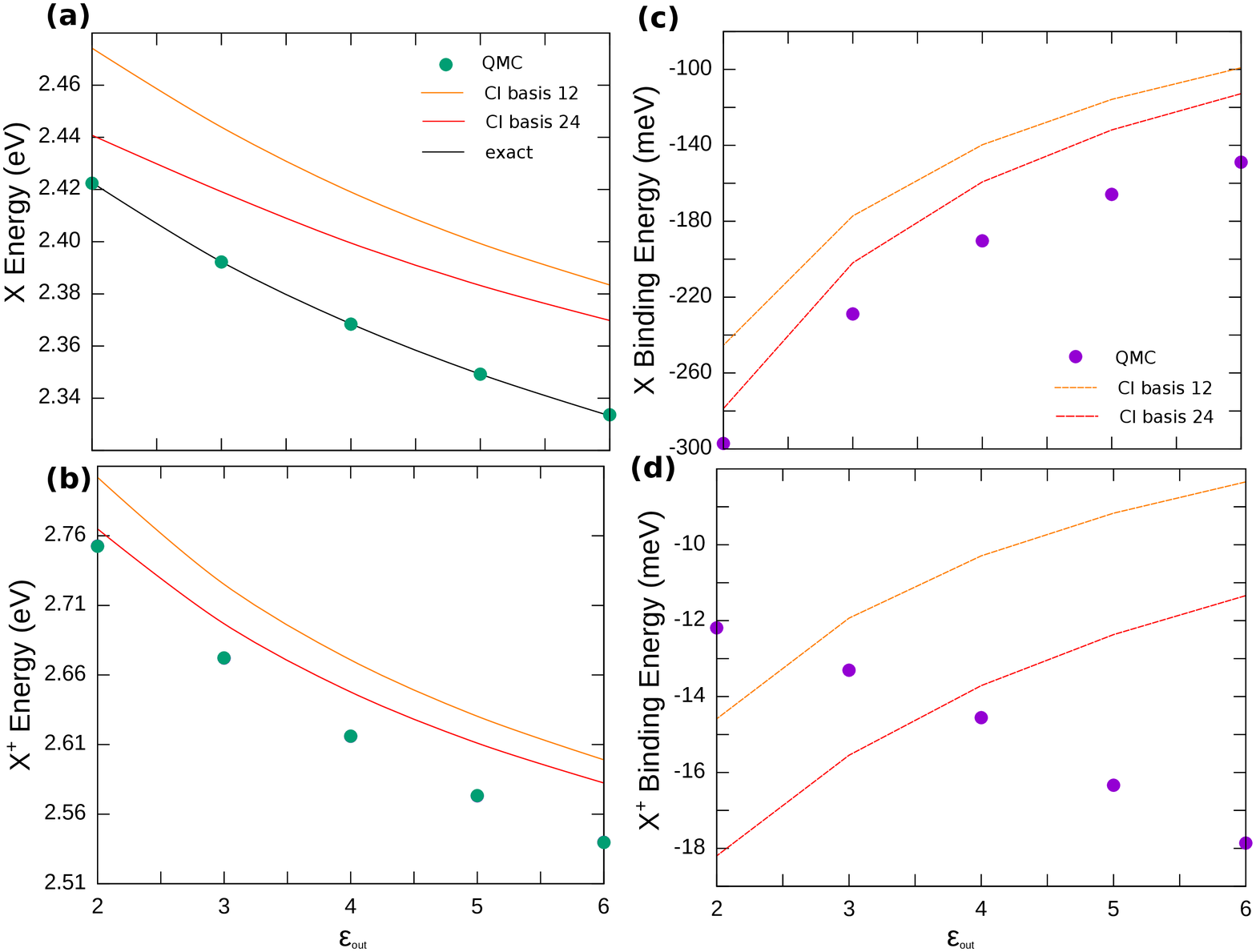}
\caption{Ground state total (a-b) and binding (c-d) energies in CdSe NPLs as a function of the outer medium dielectric constant.
(a,c) Neutral exciton. (b,d) Positive trion. Dots are VQMC calculations, with $(m,nw)=(2\cdot10^6,600)$.
Colored lines are CI calculations with a basis set built from single-particle spin-orbitals.
Black line in (a) indicates the full variational calculation.
	.}\label{fig3}
\end{center}
\end{figure}

As an illustrative calculation, Figure \ref{fig3} compares the energy of neutral and 
charged excitons in a CdSe NPL with $(L_x,L_y,L_z)=(30,10,1.4)$ nm, as a function of
the dielectric mismatch. The outer medium dielectric constant is varied from $\epsilon_{out}=2$,
which is a typical value for organic ligands passivating NPLs,\cite{AchtsteinNL}
to $\epsilon_{out}=\epsilon_{in}$, which suppresses dielectric confinement.
Fig.~\ref{fig3}(a) and (b) show the total energy of exciton and trion, respectively.
The exciton behavior is well known. With decreasing dielectric contrast, the energy
decreases due to the weakening of self-energy repulsion, which exceeds the
weakening of electron-hole Coulomb attraction.\cite{PolovitsynCM}
We observe this behavior both with VQMC calculations (dots) and with the full variational
integration of Ref.~\cite{RajadellPRB} (black line).
The same trend is found for the positive trion, which reveals that self-energy terms 
prevail over Coulomb enhancement in charged excitons too. In fact, the shift in energy is
more pronounced than for the exciton, because the larger number of particles translates
into a more relevant contribution of self-energy repulsion.

Colored lines in Fig.~\ref{fig3} show exciton and trion energies calculated using a CI method, 
as described in Ref.~\cite{MovillaJPCL}. Two basis sets are used, one built from all possible combinations
(Hartree products and, for the trion, Slater determinants) obtained from the lowest 12 electron and hole
spin-orbitals, and another from the 24 lowest ones. While the qualitative trend is consistent with
that of the QMC calculations, a deviation of few tens of meV is observed in both cases.
It is clear that CI calculations are far from convergence, and VQMC offer more accurate description.
 %of Coulomb binding energies, stabilizing both X and X$^+$.
Remarkably, this is in spite of VQMC calculations requiring CPU time of several minutes (see next),
as compared to few days in the case of the CI method.\cite{onCI}

The binding energies of exciton and trion, which are known to be sensitive to correlation energy,\cite{ShumwayPRB} 
are plot explicitly in Figs.~\ref{fig3}(c) and (d), respectively.
For the exciton, we define $E_b(X) = E(X) - E_e - E_h$, 
with $E(X)$ the total exciton energy, $E_e$ that of an independent electron and $E_h$ that of an independent hole.
That is, $E_b(X)$ is the stabilization energy as compared a to non-interacting electron and hole pair.
For the trion, it is defined as $E_b(X^+)=E(X^+)-E(X)-E_h$, where $E(X^+)$ stands for the total energy of the trion.
Then, $E_b(X^+)$ is the stabilization energy compared to one exciton plus one hole.
One can see that CI results underestimate the binding energies as compared to VQMC calculations, 
which is indicative of the latter capturing a larger amount of correlation energy. 
In the case of the trion, the improved description reveals a qualitative trend 
opposed to that predicted by CI calculations, whereby $E_b(X^+)$ is weakened for small $\varepsilon_{out}$.
This is indicative of dielectric mismatch enhancing Coulomb repulsions,
$V_c(\mathbf{r}_{h1},\mathbf{r}_{h2})$, over attractions, $V_c(\mathbf{r}_{e},\mathbf{r}_{h1(2)})$. 

\begin{figure}[t]
\begin{center}
\includegraphics[width=14cm]{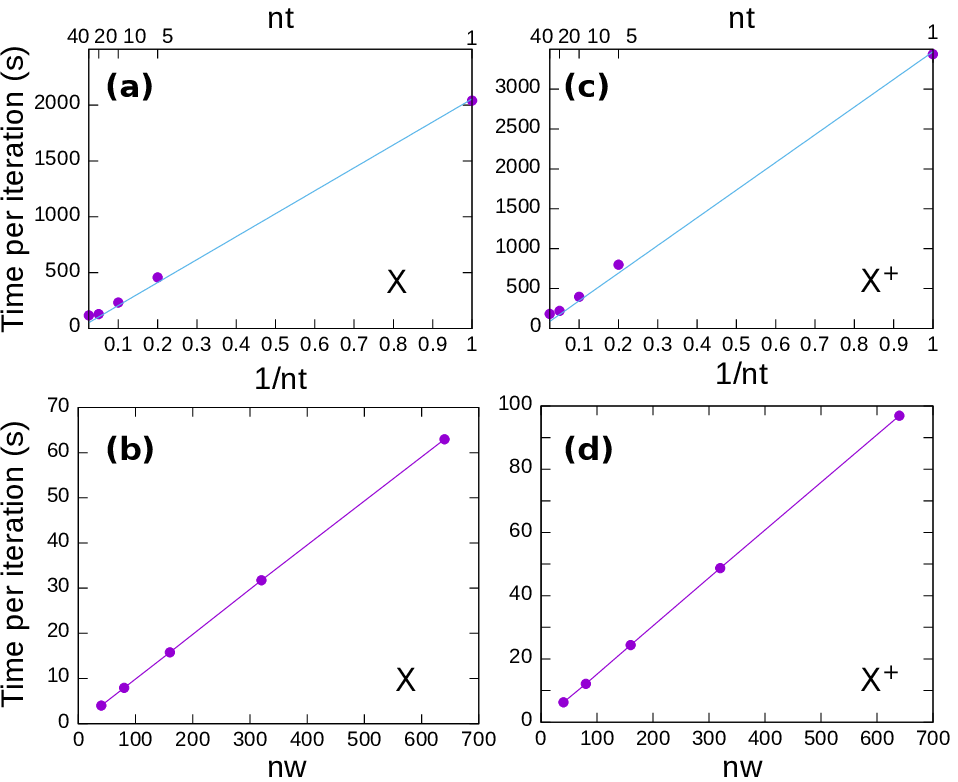}
\caption{Execution time per iteration of the Newton-Rhapson solver in the VQMC model. 
	(a) Exciton, as a function of the number of threads. 
	(b) Exciton, as a function of the number of walkers.
	(c) Trion, as a function of the number of threads. 
	(d) Trion, as a function of the number of walkers.
	In (a) and (c), ($n,nw$)=($2\cdot10^6,600$).
	In (b) and (d), $nt=40$. 
	The calculations correspond to the same NPL studied in Fig.~\ref{fig3}.
	}\label{fig4}
\end{center}
\end{figure}

Figure \ref{fig4} provides more details on the computational efficiency of the VQMC model.
As mentioned before, the calculation of random walkers can be distributed among different
threads using OpenMP directives. We benchmark the results of the parallelization in the figure.
Simulations were run on a workstation with Dual Intel Xeon E5-2660v4 core (2.2 GHz),
which admits a number of parallel threads (hereafter $nt$) of up to 40.
The code was compiled with Intel Fortran compiler and OpenMP library {\emph libomp5}.
For the neutral exciton, the time required per iteration of the Newton-Rhapson process
shows a nearly perfect scaling with $1/nt$, see Fig.~\ref{fig4}(a).
For energy convergence threshold of about 1 meV, a few iterations suffice.
Then, the total execution time is of a few minutes only.
The scaling versus number of walkers $nw$, for a fixed number of threads, 
shows a nearly linear behavior as well, see Fig.~\ref{fig4}(b).
Linear scalings hold for trions too, as shown in Figs.~\ref{fig4}(c) and (d), 
albeit the execution times increase by a factor $\sim 1.5$, 
owing to the larger number of coordinates, wave function terms and variational 
parameters compared to excitons.
The linearity of the previous plots confirms that the distribution of walkers
among threads is well balanced.  

\section{Conclusions}

We have developed \emph{Fortran} programs to calculate exciton and trion
ground state properteis in colloidal NPLs.
The codes are based on a VQMC-effective mass model which can be easily extended
to semiconductor QDs with different shapes and potentials. 
The model is conceptually simple and computationally efficient (fast execution
times, further accelerated by OpenMP parallelization, and minimal memory requirements),
and hence susceptible of being used in standard desktop computers.
By using Jastrow factors which capture short range interactions, it outperforms standard 
CI calculations in both accuracy and computational efficiency. 
It also outpeforms full variational integrations used in previous studies for
weakly confined NPLs, in that it gives access not only to neutral exciton but also 
to trion species, with only a moderate increase in execution time.

\section{Acknowledgments}
Support from MICINN project CTQ2017-83781-P and 
UJI project B2017-59 is gratefully acknowledged.

%% The Appendices part is started with the command \appendix;
%% appendix sections are then done as normal sections

%% References with bibTeX database:

\bibliographystyle{elsarticle-num}
\bibliography{vqmc}

\begin{thebibliography}{10}
\expandafter\ifx\csname url\endcsname\relax
  \def\url#1{\texttt{#1}}\fi
\expandafter\ifx\csname urlprefix\endcsname\relax\def\urlprefix{URL }\fi
\expandafter\ifx\csname href\endcsname\relax
  \def\href#1#2{#2} \def\path#1{#1}\fi

\bibitem{Pawel_book}
L.~Jacak, P.~Hawrylak, A.~Wojs, Quantum Dots, Springer, 1998.

\bibitem{Bimberg_book}
D.~Bimberg, M.~Grudmann, N.~N. Ledentsov, Quantum Dot Heterostructures, Wiley,
  1998.

\bibitem{Tapash_book}
T.~Chakraborty, Quantum Dots: A Survey of the Properties of Artificial Atoms,
  Elsevier, 1999.

\bibitem{ShumwayPRB}
J.~Shumway, A.~Franceschetti, A.~Zunger, Correlation versus mean-field
  contributions to excitons, multiexcitons, and charging energies in
  semiconductor quantum dots, Phys. Rev. B 63 (2001) 155316.

\bibitem{LhuillierACR}
E.~Lhuillier, S.~Pedetti, S.~Ithurria, B.~Nadal, H.~Heuclin, B.~Dubertret,
  Two-dimensional colloidal metal chalcogenides semiconductors: synthesis,
  spectroscopy, and applications, Accounts of Chemical Research 48 (2015)
  22--30.

\bibitem{DirollJMCc}
B.~T. Diroll, Colloidal quantum wells for optoelectronic devices, Journal of
  Materials Chemistry C (2020).

\bibitem{RichterPRM}
M.~Richter, Nanoplatelets as material system between strong confinement and
  weak confinement, Phys. Rev. Mater. 1 (2017) 016001.

\bibitem{RajadellPRB}
F.~Rajadell, J.~I. Climente, J.~Planelles, {Excitons in core-only, core-shell
  and core-crown CdSe nanoplatelets: Interplay between in-plane electron-hole
  correlation, spatial confinement, and dielectric confinement}, {Phys. Rev. B}
  {96} ({2017}) {035307}.

\bibitem{ClimentePRB}
J.~I. Climente, A.~Bertoni, G.~Goldoni, Photoluminescence spectroscopy of
  trions in quantum dots: A theoretical description, Phys. Rev. B 78 (2008)
  155316.

\bibitem{RontaniJCP}
M.~Rontani, C.~Cavazzoni, D.~Bellucci, G.~Goldoni, Full configuration
  interaction approach to the few-electron problem in artificial atoms, The
  Journal of Chemical Physics 124 (2006) 124102.

\bibitem{CorniPRB}
S.~Corni, M.~Brasken, M.~Lindberg, J.~Olsen, D.~Sundholm, Size dependence of
  the electron-hole recombination rates in semiconductor quantum dots, Phys.
  Rev. B 67 (2003) 045313.

\bibitem{BallesterJAP}
A.~Ballester, J.~Movilla, J.~Escart{\'\i}n, M.~Pi, J.~Planelles, Configuration
  interaction approach to fermi liquid--wigner crystal mixed phases in
  semiconductor nanodumbbells, Journal of Applied Physics 112 (2012) 024311.

\bibitem{RontaniJPB}
M.~Rontani, G.~Eriksson, S.~{\AA}berg, S.~Reimann, On the renormalization of
  contact interactions for the configuration-interaction method in
  two-dimensions, Journal of Physics B 50 (2017) 065301.

\bibitem{CeperleyPRB}
D.~Ceperley, G.~V. Chester, M.~Kalos, Monte carlo simulation of a many-fermion
  study, Physical Review B 16 (1977) 3081.

\bibitem{NeedsJPCM}
R.~J. Needs, M.~D. Towler, N.~D. Drummond, P.~L. R{\'\i}os, Continuum
  variational and diffusion quantum monte carlo calculations, Journal of
  Physics: Condensed Matter 22 (2009) 023201.

\bibitem{HuPRL}
Y.~Hu, S.~W. Koch, M.~Lindberg, N.~Peyghambarian, E.~Pollock, F.~F. Abraham,
  Biexcitons in semiconductor quantum dots, Phys. Rev. Lett. 64 (1990) 1805.

\bibitem{WangJAP}
E.~Wang, Y.~Zhou, C.~Ting, J.~Zhang, T.~Pang, C.~Chen, Excitons in spatially
  separated electron--hole systems: A quantum monte carlo study, J. Appl. Phys.
  78 (1995) 7099--7102.

\bibitem{WimmerPRB}
M.~Wimmer, S.~Nair, J.~Shumway, Biexciton recombination rates in self-assembled
  quantum dots, Phys. Rev. B 73~(16) (2006) 165305.

\bibitem{LeeNL}
D.~Lee, J.~L. DuBois, Y.~Kanai, Importance of excitonic effect in charge
  separation at quantum-dot/organic interface: First-principles many-body
  calculations, Nano Lett. 14 (2014) 6884--6888.

\bibitem{McDonaldPRB}
P.~G. McDonald, E.~J. Tyrrell, J.~Shumway, J.~M. Smith, I.~Galbraith, Tuning
  biexciton binding and antibinding in core/shell quantum dots, Physical Review
  B 86 (2012) 125310.

\bibitem{casino}
\href{https://vallico.net/casinoqmc/}{[link]}.
\newline\urlprefix\url{https://vallico.net/casinoqmc/}

\bibitem{qmcpack}
\href{https://qmcpack.org/}{[link]}.
\newline\urlprefix\url{https://qmcpack.org/}

\bibitem{WilliamsonPRL}
A.~J. Williamson, J.~C. Grossman, R.~Q. Hood, A.~Puzder, G.~Galli, Quantum
  monte carlo calculations of nanostructure optical gaps: Application to
  silicon quantum dots, Phys. Rev. Lett. 89 (2002) 196803.

\bibitem{KumagaiPRB}
M.~Kumagai, T.~Takagahara, {Excitonic and Nonlinear-Optical Properties of
  Dielectric Quantum-Well Structures}, Phys. Rev. B. 40 (1989) 12359--12381.

\bibitem{Bastard_book}
G.~Bastard, Wave mechanics applied to semiconductor heterostructures, Les
  Editions de Physique, 1988.

\bibitem{KatoCPAM}
T.~Kato, On the eigenfunctions of many-particle systems in quantum mechanics,
  Communications on Pure and Applied Mathematics 10~(2) (1957) 151--177.

\bibitem{PlanellesTCA}
J.~Planelles, Simple correlated wave-function for excitons in 0d, quasi-1d and
  quasi-2d quantum dots, Theoretical Chemistry Accounts 136~(7) (2017) 81.

\bibitem{SergeevPSS}
R.~Sergeev, R.~Suris, Ground-state energy of x- and x+ trions in a
  two-dimensional quantum well at an arbitrary mass ratio, Physics of the Solid
  State 43 (2001) 746--751.

\bibitem{LinJCP}
X.~Lin, H.~Zhang, A.~M. Rappe, Optimization of quantum monte carlo wave
  functions using analytical energy derivatives, The Journal of Chemical
  Physics 112 (2000) 2650--2654.

\bibitem{UmrigarPRL}
C.~J. Umrigar, C.~Filippi, Energy and variance optimization of many-body wave
  functions, Phys. Rev. Lett. 94 (2005) 150201.

\bibitem{general}
The energy in Eq.~(12) is replaced by the magnitude of interest.

\bibitem{ders}
Obtaining accurate estimates of the derivatives in Eqs.(18-20), rather than
  local energy itself, is the most demanding step in the calculation.

\bibitem{AchtsteinNL}
A.~W. Achtstein, A.~Schliwa, A.~Prudnikau, M.~Hardzei, M.~V. Artemyev,
  C.~Thomsen, U.~Woggon, Electronic structure and exciton--phonon interaction
  in two-dimensional colloidal cdse nanosheets, Nano Lett. 12 (2012)
  3151--3157.

\bibitem{onself}
The self-energy potential in Eq.~(22) does not depend on in-plane coordinates.
  To avoid the associated variance, we integrate it separately and add the
  result to the local energy.

\bibitem{Sadao_book}
S.~Adachi, Handbook on Physical Properties of Semiconductors, vol.3, Kluwer
  Academics, 2004.

\bibitem{PolovitsynCM}
A.~Polovitsyn, Z.~Dang, J.~L. Movilla, B.~Mart{\'\i}n-Garc{\'\i}a, A.~H. Khan,
  G.~H. Bertrand, R.~Brescia, I.~Moreels, Synthesis of air-stable cdse/zns
  core--shell nanoplatelets with tunable emission wavelength, Chem. Mater. 29
  (2017) 5671--5680.

\bibitem{MovillaJPCL}
J.~L. Movilla, J.~Planelles, J.~I. Climente, {Dielectric Confinement Enables
  Molecular Coupling in Stacked Colloidal Nanoplatelets}, {J. Phys. Chem.
  Lett.} {11} ({2020}) {3294--3300}.

\bibitem{onCI}
The calculation of a high number of six-dimensional Coulomb integrals is the
  computational bottleneck in the CI method.

\end{thebibliography}

%% Authors are advised to submit their bibtex database files. They are
%% requested to list a bibtex style file in the manuscript if they do
%% not want to use elsarticle-num.bst.

\end{document}